\documentclass[aps,prl,preprint,groupedaddress]{revtex4-1}
\usepackage{graphicx}% Include figure files
\usepackage{dcolumn}% Align table columns on decimal point
\usepackage{bm}% bold math
\usepackage{color}

\usepackage{comment}
\usepackage[geometry]{ifsym}
\usepackage{booktabs}
\usepackage[geometry]{ifsym}
\usepackage{amsthm}
\usepackage{amsmath}

\theoremstyle{remark}

\usepackage{amssymb}
\usepackage{booktabs}
\usepackage{amsthm}

\usepackage{enumerate}
\usepackage{multirow}
\begin{document}

% Use the \preprint command to place your local institutional report
% number in the upper righthand corner of the title page in preprint mode.
% Multiple \preprint commands are allowed.
% Use the 'preprintnumbers' class option to override journal defaults
% to display numbers if necessary
%\preprint{}

%Title of paper
\title{A Cascade of Volterra-Operator BBP Transitions  in a Correlated Wigner Matrix
}

% repeat the \author .. \affiliation  etc. as needed
% \email, \thanks, \homepage, \altaffiliation all apply to the current
% author. Explanatory text should go in the []'s, actual e-mail
% address or url should go in the {}'s for \email and \homepage.
% Please use the appropriate macro foreach each type of information

% \affiliation command applies to all authors since the last
% \affiliation command. The \affiliation command should follow the
% other information
% \affiliation can be followed by \email, \homepage, \thanks as well.

\author{Masato Hisakado}
\email{hisakadom@yahoo.co.jp} 
\affiliation{
*Kanazwa university, Kakumamachi, Kanazwa, Ishikawa 920-1192, Japan} 
\begin{comment}
\author{Kazuaki Nakayama}
\email{tkaneko@icu.ac.jp}
\affiliation{
\dag 
Shinsyu university\\
Asahi 3-1-1, Matsumoto, Nagano 390-8621, Japan}
\end{comment}
%\end{}

%Collaboration name if desired (requires use of superscriptaddress
%option in \documentclass). \noaffiliation is required (may also be
%used with the \author command).
%\collaboration can be followed by \email, \homepage, \thanks as well.
%\collaboration{}
%\noaffiliation

\date{\today}

\begin{abstract}
We study a Wigner-type random matrix in which the off-diagonal correlation between entries is generated by a random factor shared among all entries in a given row and column, with the coupling strength held fixed as the matrix size grows. Although the bulk spectral moments remain those of the pure semicircle law, we show that the underlying correlation matrix decomposes into a vanishing bulk together with a countable family of outlier eigenvalues that, at fixed rank $k$, converge to the singular values of a compact Volterra (cumulative-sum) integral operator -- obtained in closed form via the classical Karhunen--Lo\`eve expansion of Brownian motion and confirmed numerically to better than one percent across the top twenty such values. Each singular value drives an independent Baik--Ben Arous--P\'{e}ch\'{e} (BBP) transition as the coupling strength increases, producing an evenly spaced, discrete hierarchy of critical points -- rather than a single transition -- at each of which one further eigenvalue detaches from the semicircle edge, in close agreement with direct diagonalization. We show that this mechanism generalizes to a broader family of correlation structures, with the critical hierarchy in every case set by the spectrum of an associated compact integral operator.

\hspace{0cm}
\vspace{1cm}
%PACS numbers: {02.50.Ga, 05.70.Fh, 89.65.Gh, 87.23.Kg} 
\end{abstract}

%02.50.Ga, 05.70.Fh, 89.65.Gh, 87.23.Kg
%\pacs{02.50.Ga, 05.70.Fh, 89.65.Gh, 87.23.Kg} 
%\keywords{Pitman sampling formula, Kirman's Ant colony model, Ising model}

%\maketitle must follow title, authors, abstract, \pacs, and \keywords
\maketitle

\section{I. Introduction}

Random matrix theory (RMT) provides a universal framework for describing the spectral
behavior of large complex systems~\cite{Meh}. Wigner matrices with independent,
identically distributed entries converge, after appropriate normalization, to the semicircle
law, while Wishart matrices converge to the Marchenko--Pastur distribution; deviations from
these classical laws, arising from correlations, heavy tails, or other structural deformations,
have been studied extensively across physics, mathematics, and applications ranging from
nuclear spectra to network theory and finance~\cite{Pot,Pot3,Pot2,Fra,RM,AI}.

Among such deformations, the Baik--Ben Arous--P\'{e}ch\'{e} (BBP) transition~\cite{BBP2005}
occupies a distinguished place: a Wigner or Wishart matrix perturbed by a finite-rank
deterministic spike undergoes a sharp phase transition, as the spike strength increases past
a critical threshold, from Tracy--Widom edge statistics to a detached eigenvalue with Gaussian
fluctuations. This result and its refinements~\cite{FeralPeche2007,BloemendalVirag2013,BaikSilverstein2006,Paul2007}
have become a standard tool for detecting low-rank signals in high-dimensional data.

In this paper we study a Wigner-type matrix in which the perturbation is not a finite-rank
deterministic spike but is instead generated by a random factor $Z_j$ shared among all
entries in a given row and column -- a natural model, for example, of a latent common
driver contaminating a correlation matrix built from finite time series. Unlike a single
global common factor, which produces a trivial rank-one perturbation, the specific
sharing pattern considered here produces a perturbation matrix that is full rank at every
finite $N$, yet whose relevant large-$N$ behavior is governed entirely by the spectrum of
an associated compact integral operator -- the Volterra (cumulative-sum) operator, whose
singular values follow from the classical Karhunen--Lo\`eve expansion of Brownian motion.
This structure elevates the BBP transition from a single critical point to a discrete,
countably infinite hierarchy of critical points, one for each singular value of the governing
operator, and connects the resulting cascade's density directly to operator-theoretic
properties -- compactness and the trace-class property -- of the underlying kernel.
The spectrum of the compact operator quantizes the phase transition of the random matrix.

The remainder of this paper is organized as follows. In Section II, we introduce the
correlated Wigner matrix and its bulk spectral moments. In Section III, we decompose the
correlation matrix and identify its bulk-plus-spikes structure. In Section IV, we relate
the spikes to the spectrum of the Volterra operator. In Section V, we derive the resulting
cascade of BBP transitions. Section VI discusses the generalization of this mechanism to
other kernels and its relation to spiked Wishart models. Conclusions are presented in
Section VII.

\section{II.  Creation of the Wigner matrix with  correlation}

\subsection{A. Definition}
In this section,  we  introduce  the Wigner     matrix with  correlation. 
In the matrix form, we can write  the $N\times N$ matrix
form
\begin{eqnarray}
S&=&S_{ij}=
\left(
    \begin{array}{ccccc}
     A_{11} &  A_{12}&A_{13}  &\cdots& A_{1N} \\
   A_{12}    &  A_{22} & A_{23}&\cdots& A_{2N} \\
   A_{13} &A_{23}& A_{33}& \cdots           & A_{3N} \\
 \vdots & \vdots & \ddots &\cdots &\vdots\\
    A_{1N}& A_{2N}& A_{3N} &  \cdots &A_{NN} \\
    \end{array}
  \right),
  \label{EX}
\end{eqnarray}
where
\begin{equation}
  A_{ij}=\frac{bZ_j}{\sqrt{N}}+\sigma \eta_{ij},
  \label{2}
\end{equation}
where
$j=1,2, \cdots, N$ and $i=1,\cdots, j-1$.
$Z_{j}$ and $\eta_{ij}$ are i.i.d. and
$b$, $\sigma$ are constants.

This construction ensures that the matrix elements inherit the  correlation structure
Hence, 
\begin{equation}
\mbox{Cov}(A_{ij},A_{i'j})=\frac{b^2}{N},
\end{equation}
 and 
\begin{equation}
\mbox{Cov}(A_{ij},A_{i'j'})=0,
\end{equation}
when $j\neq j'$.

% =====================================================================
% Continuation draft: Section III onward
% Notation carried over from Section II (eqs. (1)-(4)):
%   S = (S_ij), A_ij = b Z_{max(i,j)} / sqrt(N) + sigma eta_ij  (i<j)
%   Z_j, eta_ij : i.i.d., mean 0, variance 1
%   b, sigma : constants (b fixed, does not scale with N)
% New notation introduced below:
%   H = S / sqrt(N)              (standard Wigner normalization)
%   M_ij = Z_{max(i,j)}, M_ii=0  (so that S = (b/sqrt(N)) M + sigma eta)
%   U    : strictly upper triangular all-ones matrix, U_{ij}=1 (i<j)
%   D    = diag(Z_1,...,Z_N)
% =====================================================================
 
\subsection{B. Bulk moments and the standard normalization}
 
It is convenient to work with the standard Wigner normalization
\begin{equation}
H \equiv \frac{S}{\sqrt N}, \qquad
H_{ij} = \frac{b}{N} Z_{\max(i,j)} + \frac{\sigma}{\sqrt N}\eta_{ij} .
\label{eq:Hdef}
\end{equation}
Writing $M_{ij} \equiv Z_{\max(i,j)}$ for $i\neq j$ and $M_{ii}=Z_i$, Eq.~\eqref{eq:Hdef} takes the compact form
\begin{equation}
H = \frac{b}{N}M + \frac{\sigma}{\sqrt N}\eta .
\label{eq:Hcompact}
\end{equation}
Since $Z_j$ are independent of one another (no temporal correlation is assumed here, in contrast to Ref.~\cite{HisakadoKaneko2025}), a direct computation of the second and fourth moments gives, to leading order in $N$,
\begin{align}
\frac{1}{N}\,\mathbb E\,(\mathrm{Tr}\,H^2) &\to \sigma^2, \label{eq:m2}\\
\frac{1}{N}\,\mathbb E\,(\mathrm{Tr}\,H^4) &\to 2\sigma^4, \label{eq:m4}
\end{align}
independently of $b$. Both moments coincide with those of the semicircle law of radius $2\sigma$. This might suggest that the parameter $b$ plays no role in the $N\to\infty$ limit. We show in the following sections that this conclusion is correct for the \emph{bulk} of the spectrum but fails at the \emph{edge}: for $b$ larger than a critical value $b_c$, an isolated eigenvalue detaches from the bulk.
 
\section{III Spectral decomposition of $M$: bulk and spikes}
\label{sec:decomposition}
 
The matrix $M$ admits the exact decomposition
\begin{equation}
M = UD + (UD)^\top -D,
\label{eq:Mdecomp}
\end{equation}
where $U$ is the upper triangular all-ones matrix, $U_{ij}=1$ for $i\leq j$ and $U_{ij}=0$ otherwise, and $D=\mathrm{diag}(Z_1,\dots,Z_N)$. Indeed, $(UD)_{ij}=Z_j$ for $i < j$, so that $\big(UD+(UD)^\top\big)_{ij}=Z_j$ for $i<j$ and $=Z_i$ for $i>j$, $=2 Z_i$ for $i=j$, and  i.e.\ $M_{ij}=Z_{\max(i,j)}$ as required.
Note that the third term of Eq.(\ref{eq:Mdecomp}) is for  diagonal elements.
 
\begin{figure}[h]
\centering
\includegraphics[width=\columnwidth]{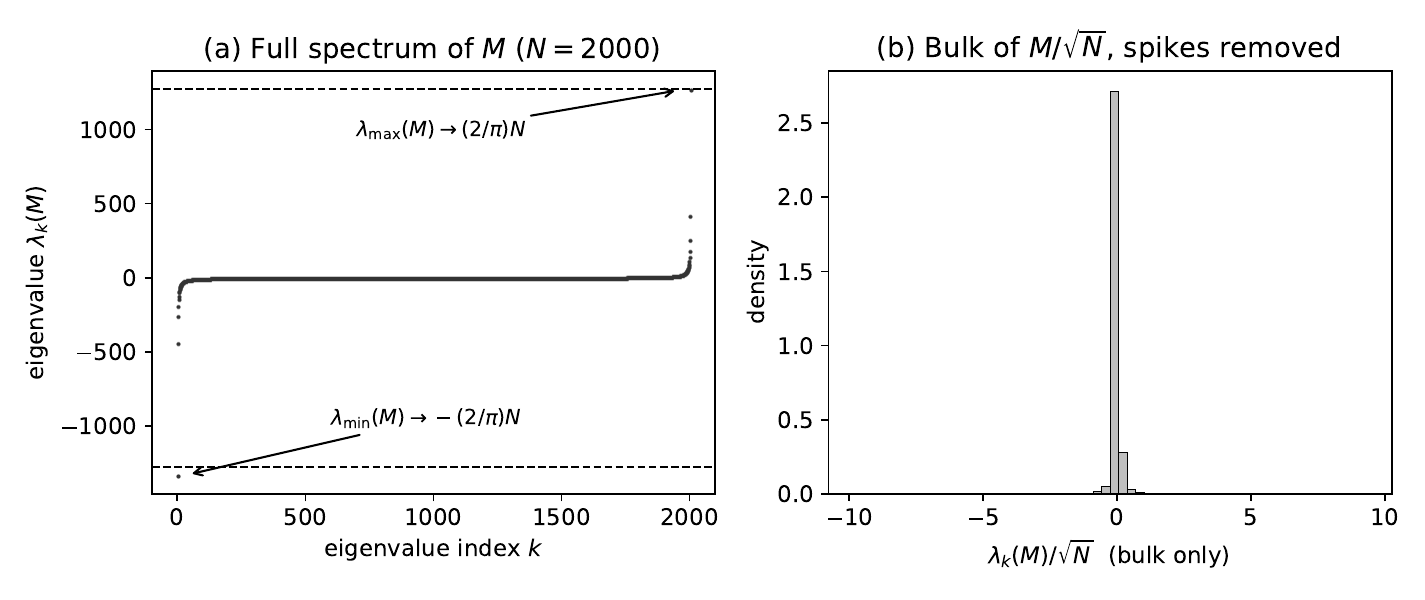}
\caption{(a) Full eigenvalue spectrum of $M$ ($N=2000$, one realization), sorted eigenvalue $\lambda_k(M)$ vs.\ rank $k$. Two eigenvalues detach from the bulk near $\pm(2/\pi)N$ (dashed lines). (b) Histogram of the remaining $N-2$ bulk eigenvalues, rescaled by $\sqrt N$, after removing the two outliers.}
\label{fig:spectrum_M}
\end{figure}

Numerical diagonalization of $M$ (Fig.~\ref{fig:spectrum_M}) reveals a clean separation of scales in its spectrum:
\begin{itemize}
\item $N-2$ eigenvalues form a \emph{bulk} of order $O(\sqrt N)$ (equivalently, $O(1)$ after dividing by $\sqrt N$);
\item two eigenvalues, one near the top and one near the bottom of the spectrum, are isolated from the bulk and scale as $O(N)$.
\end{itemize}

Table~\ref{tab:spikeM} lists the largest eigenvalue of $M$ obtained from independent realizations at several values of $N$; the ratio $\lambda_{\max}(M)/N$ converges rapidly to a constant,
\begin{equation}
\frac{\lambda_{\max}(M)}{N} \ \longrightarrow\ c, \qquad c \approx 0.636,
\label{eq:cnumeric}
\end{equation}
with the corresponding bottom eigenvalue converging to $-cN$ by the symmetry $Z_j\to -Z_j$. We identify the constant $c$ analytically in Sec.~IV and find $c=2/\pi$, in agreement with Eq.~\eqref{eq:cnumeric}; this identification is corroborated independently in Sec. IV using the fourth moment $\mathrm{Tr}(M^4)$.
 
\begin{table}[h]
\centering
\begin{tabular}{c c c}
\hline\hline
$N$ & $\lambda_{\max}(M)/N$ & $\lambda_{\max}(M)/N - 2/\pi$ \\
\hline
1000 & 0.639 & $+0.003$ \\
2000 & 0.6365 & $-0.0001$ \\
4000 & 0.6386 & $+0.0020$ \\
8000 & 0.6360 & $-0.0006$ \\
\hline\hline
\end{tabular}
\caption{Largest eigenvalue of $M$, rescaled by $N$, averaged over independent realizations, compared with $2/\pi = 0.63662\ldots$.}
\label{tab:spikeM}
\end{table}
 
Given Eq.~\eqref{eq:cnumeric}, the two outliers of $M$ contribute to $H$ [Eq.~\eqref{eq:Hcompact}] a term of order
\begin{equation}
\frac{b}{N}\cdot cN = cb = O(1),
\end{equation}
which does \emph{not} vanish as $N\to\infty$, unlike the bulk contribution of $M$ [which is suppressed by an extra factor $1/\sqrt N$ relative to the noise term $\sigma\eta/\sqrt N$, consistent with Eqs.~\eqref{eq:m2}--\eqref{eq:m4}]. The problem of the extreme eigenvalues of $H$ therefore reduces, at leading order in $N$, to a Wigner matrix of noise scale $\sigma$ additively perturbed by two eigenvalues of fixed magnitude $\pm cb$: a Baik--Ben Arous--P\'{e}ch\'{e} (BBP) deformation problem \cite{BBP2005}, which we analyze in Sec.~V after identifying the constant $c$.
%%%%%%%%%%%%%%%%%%%%%%%%%
\section{IV The Volterra operator: full spectrum and the emergence of a transition hierarchy}
\label{sec:volterra}

\subsection{A. Continuum correspondence}

Let $U$ denote the upper triangular all-ones matrix including the diagonal, $U_{ij}=1$ for $i\le j$, so that $M=UD+(UD)^\top-D$ (Sec.~III). For a smooth test vector $x_j=\varphi(j/N)$, $\varphi\in L^2[0,1]$, and $t=i/N$,
\begin{equation}
(Ux)_i = \sum_{j\ge i} \varphi(j/N) \ \approx\ N\int_t^1 \varphi(s)\,ds ,
\label{eq:Ucontinuum}
\end{equation}
which identifies $U$, in the large-$N$ limit and on sufficiently smooth test vectors, with $N$ times the operator
\begin{equation}
(A\varphi)(t) \equiv \int_t^1 \varphi(s)\,ds
\label{eq:Adef}
\end{equation}
on $L^2[0,1]$ -- the Volterra operator run backward from $t=1$. (This differs from the forward Volterra operator of Sec.~III's original presentation only by the reflection $t\to1-t$, which leaves the operator norm and singular values unchanged.) $A$ is a Hilbert--Schmidt integral operator, hence compact.
%%%%%%%%%%%%%%%%%%

\paragraph{The complementary (forward) convention.} Reversing the triangular convention, $L\equiv U^\top$ ($L_{ij}=1$ for $i\geq j$), gives $M^{(L)}=LD+(LD)^\top-D$ with $M^{(L)}_{ij}=Z_{\min(i,j)}$ in place of $M_{ij}=Z_{\max(i,j)}$: the roles of $i$ and $j$ in the defining recursion are exchanged, replacing the shared factor at the \emph{larger} index by the shared factor at the \emph{smaller} one. By the same argument as above, $L$ corresponds in the continuum limit to the forward Volterra operator $(A_{\mathrm{fwd}}\varphi)(t)=\int_0^t\varphi(s)\,ds$ -- the operator one might naively have associated with $U$ before accounting for the direction of the sum in Eq.~\eqref{eq:Ucontinuum}. Since $A_{\mathrm{fwd}}$ and $A$ are related by the reflection $t\mapsto1-t$, they share the same singular values, $\sigma_k(A_{\mathrm{fwd}})=\sigma_k(A)$; direct diagonalization of $M^{(L)}$ confirms that its eigenvalue distribution is statistically indistinguishable from that of $M$ (Table~\ref{tab:k20}), consistent with the reflection symmetry $i\mapsto N+1-i$ relating the two constructions. All subsequent results of this paper -- the spike constant $2/\pi$, the full spectrum~\eqref{eq:fullspectrum}, and the transition hierarchy of Sec.~V -- are therefore identical whether one adopts the $Z_{\max(i,j)}$ or the $Z_{\min(i,j)}$ convention; we retain the former throughout for consistency with Eq.(~\ref{2}) of Sec.~II.

%%%%%%%%%%%%%%%%%%%%%%%%%%%
\subsection{B. Full singular value spectrum}
The operator $A^*A$ has integral kernel
\begin{equation}
(A^*A f)(s) = \int_0^1 \min(s,t)\, f(t)\, dt,
\label{eq:AstarA}
\end{equation}
which is precisely the covariance kernel of standard Brownian motion, $\mathrm{Cov}(B_s,B_t)=\min(s,t)$. The eigenvalue problem $A^*A f = \mu f$ is the classical Karhunen--Lo\`eve problem for Brownian motion. Differentiating Eq.~\eqref{eq:AstarA} twice with respect to $s$ yields the Sturm--Liouville problem
\begin{equation}
f''(s) = -\frac{1}{\mu} f(s), \qquad f(0)=0,\quad f'(1)=0,
\label{eq:SL}
\end{equation}
with boundary conditions inherited from $\min(0,t)=0$ and $\partial_s\min(s,t)|_{s=1}=0$ for $t<1$. The solutions are $f_k(s)=\sin\!\big((k-\tfrac12)\pi s\big)$, $k=1,2,\dots$, with eigenvalues
\begin{equation}
\mu_k = \frac{1}{\pi^2(k-\tfrac12)^2}.
\label{eq:muk}
\end{equation}
\begin{comment}
The singular values of $A$ are $\sigma_k(A)=\sqrt{\mu_k}=1/\big(\pi(k-\tfrac12)\big)$, and the largest one is
\begin{equation}
\boxed{\ \|A\|_{\mathrm{op}} = \sigma_1(A) = \frac{2}{\pi}\ }.
\label{eq:Anorm}
\end{equation}
\end{comment}
Equation~\eqref{eq:SL} gives the \emph{complete} spectrum
\begin{equation}
\sigma_k(A) = \frac{1}{\pi\big(k-\tfrac12\big)}, \qquad k=1,2,3,\dots.
\label{eq:fullspectrum}
\end{equation}
The largest value is
\[
\boxed{\|A\|_{\mathrm{op}}=\sigma_1(A)=2/\pi}.
\]
Because $A$ is compact, this spectrum is discrete and accumulates only at $0$; because
\begin{equation}
\sum_k\sigma_k(A)^2 = \sum_k\frac{1}{\pi^2(k-\tfrac12)^2} = \frac12 < \infty, \qquad
\sum_k\sigma_k(A) = \sum_k\frac1{\pi(k-\tfrac12)} = \infty,
\label{eq:HSnottraceclass}
\end{equation}
$A$ is Hilbert--Schmidt but not trace class. Equation~\eqref{eq:HSnottraceclass} may be verified independently via the classical Fredholm determinant identity
\begin{equation}
\det\big(I-x^2A^*A\big) = \prod_{k=1}^\infty\left(1-\frac{x^2}{\pi^2(k-\tfrac12)^2}\right) = \cos x,
\label{eq:fredholm}
\end{equation}
an elementary consequence of Euler's product formula for cosine; Eq.~\eqref{eq:fredholm} in particular gives $\mathrm{Tr}(A^*A)=\sum_k\sigma_k(A)^2=1/2$ directly from the Taylor coefficient of $x^2$, matching $\int_0^1\min(t,t)\,dt=\int_0^1t\,dt=1/2$.

\subsection{C. The full edge spectrum of $M$ matches $\{\sigma_k(A)\}$}

Table~\ref{tab:k20} reports the twenty largest eigenvalues of $M/N$ (in absolute value; by the symmetry $Z_j\to-Z_j$ the twenty smallest are their negatives), averaged over six independent realizations at $N=8000$, using the natural diagonal convention $M_{ii}=Z_i$. Agreement with the prediction~\eqref{eq:fullspectrum} holds to within $1\%$ across all twenty values, including near $k=20$ where $\sigma_{20}(A)\approx0.0163$ is only about $1.5$ times the bulk edge scale $1/\sqrt N\approx0.0112$.

\begin{table}[h]
\centering
\begin{tabular}{c c c c}
\hline\hline
$k$ & $\sigma_k(A)$ & $\lambda_k(M)/N$ (observed) & ratio \\
\hline
1  & 0.6366 & 0.6402 & 1.006 \\
2  & 0.2122 & 0.2139 & 1.008 \\
3  & 0.1273 & 0.1284 & 1.009 \\
5  & 0.0707 & 0.0713 & 1.008 \\
10 & 0.0335 & 0.0337 & 1.007 \\
15 & 0.0220 & 0.0221 & 1.008 \\
20 & 0.0163 & 0.0164 & 1.005 \\
\hline\hline
\end{tabular}
\caption{Predicted vs.\ observed eigenvalues of $M/N$, $N=8000$, six realizations, natural diagonal convention $M_{ii}=Z_i$. Full $k=1,\dots,20$ list available on request; all twenty ratios lie in $[0.999,1.013]$.}
\label{tab:k20}
\end{table}

This indicates that it is not merely the top eigenvalue but the \emph{entire edge spectrum} of $M/N$ that converges to the full singular value spectrum of $A$,
\begin{equation}
\lim_{N\to\infty}\frac{\lambda_k(M)}{N} = \sigma_k(A), \qquad k=1,2,3,\dots\ \text{(fixed)},
\label{eq:fullconvergence}
\end{equation}
with the bulk of the remaining $N-O(1)$ eigenvalues vanishing as $O(1/\sqrt N)$.

\subsection{D. Status of Eq.~\eqref{eq:fullconvergence}}

As with the single-spike case discussed in earlier work on this model, we do not provide a rigorous proof of Eq.~\eqref{eq:fullconvergence}. A fixed (disorder-independent) trial vector in the Rayleigh quotient $x^\top Mx$ -- for instance the deterministic singular function of $A$ itself -- yields only an $O(\sqrt N)$ value, not the $O(N)$ required; direct numerical evaluation confirms this. The $O(N)$ scaling is achieved only by an eigenvector adapted to the specific realization of $(Z_1,\dots,Z_N)$, placing the rigorous justification of Eq.~\eqref{eq:fullconvergence} in the domain of supremum theory for Gaussian processes indexed by the sphere (generic chaining, Sudakov--Fernique comparison) rather than elementary variational bounds. We leave a complete proof for future work and rely here on the numerical evidence of Table~\ref{tab:k20}, together with independent confirmation via the fourth trace moment reported earlier.

\subsection{E. Independent verification via higher trace moments}
 
Equation~\eqref{eq:fullconvergence} can be tested collectively, rather than eigenvalue by eigenvalue, via the trace moments of $M$. Since the bulk contributes $O(N)\times O(N^{-k})=O(N^{1-k})$ to $\mathrm{Tr}(M^{2k})/N^{2k}$, this contribution vanishes for $k\ge2$ and the moment is governed entirely by the discrete spectrum~\eqref{eq:fullspectrum}:
\begin{equation}
\lim_{N\to\infty}\frac{\mathrm{Tr}(M^{2k})}{N^{2k}} = 2\sum_{j=1}^\infty\sigma_j(A)^{2k} \equiv 2\,p_k, \qquad k\ge2.
\label{eq:momentprediction2}
\end{equation}
%%%%%%%%%%%%
The power sums $p_n=\sum_k\sigma_k(A)^{2n}$ admit a fully explicit closed form for every $n$, not merely the first few values obtainable from the Taylor expansion of $\log\cos x$. Since $\sigma_k(A)=1/(\pi(k-\tfrac12))$,
\begin{equation}
p_n = \frac{1}{\pi^{2n}}\sum_{k=1}^\infty\frac{1}{(k-\tfrac12)^{2n}} = \frac{2^{2n}}{\pi^{2n}}\!\!\sum_{m\ \mathrm{odd}}\frac{1}{m^{2n}} = \frac{(2^{2n}-1)\,\zeta(2n)}{\pi^{2n}},
\label{eq:pnzeta}
\end{equation}
using the standard identity $\sum_{m\ \mathrm{odd}} m^{-s} = (1-2^{-s})\zeta(s)$. Substituting the classical evaluation $\zeta(2n) = (-1)^{n+1} B_{2n}(2\pi)^{2n}/[2(2n)!]$ in terms of Bernoulli numbers $B_{2n}$ renders Eq.~\eqref{eq:pnzeta} fully elementary,
\begin{equation}
p_n = \frac{(-1)^{n+1}\,(2^{2n}-1)\,2^{2n-1}\,B_{2n}}{(2n)!}.
\label{eq:pnbernoulli}
\end{equation}
Equation~\eqref{eq:pnbernoulli} reproduces $p_1=1/2$, $p_2=1/6$, $p_3=1/15$, $p_4=17/630$ and gives $p_n$ in closed form for arbitrary $n$, with no need for term-by-term Taylor expansion.
%%%%%%%%%%%%%%%
Table~\ref{tab:highmoments} compares these predictions with direct numerical evaluation of $\mathrm{Tr}(M^{2k})/N^{2k}$ for $k=2,3,4$; unlike Table~\ref{tab:k20}, this test does not isolate individual eigenvalues but is sensitive to the \emph{entire} spectrum~\eqref{eq:fullspectrum} simultaneously, providing an independent check of Eq.~\eqref{eq:fullconvergence} of a qualitatively different kind.
 
\begin{table}[h]
\centering
\begin{tabular}{c c c}
\hline\hline
$k$ & observed $\mathrm{Tr}(M^{2k})/N^{2k}$ ($N=2000$) & predicted $2p_k$ \\
\hline
2 & $0.332\pm0.026$ & $0.333$ \\
3 & $0.133\pm0.016$ & $0.133$ \\
4 & $0.054\pm0.008$ & $0.054$ \\
\hline\hline
\end{tabular}
\caption{Higher trace moments of $M$ at $N=2000$ (six realizations), compared with the exact predictions $2p_k=2\sum_j\sigma_j(A)^{2k}$ obtained from the Fredholm determinant identity~\eqref{eq:fredholm}.}
\label{tab:highmoments}
\end{table}

%%%%%%%%%%%%%%%%%%%%%%%
\section{V A cascade of BBP transitions}
\label{sec:bbp}
%%%%%%%%%%%%%%%

\subsection{A. Reduction to a rank-one deformation problem}

By Sec.~IV, the top eigenvalue problem for $H$ [Eq.~\eqref{eq:Hcompact}] reduces, at leading order in $N$, to
\begin{equation}
H \ \simeq\ \frac{\sigma}{\sqrt N}\,\eta \ +\ \theta\, u u^\top, \qquad \theta \equiv \frac{2b}{\pi},
\label{eq:bbpform}
\end{equation}
where $u$ is the (unit-norm) top eigenvector of $M$ and $\sigma\eta/\sqrt N$ is a standard GOE-normalized Wigner matrix with semicircular limiting spectral density of radius $2\sigma$. Equation~\eqref{eq:bbpform} is precisely the setting of the Baik--Ben Arous--P\'{e}ch\'{e} (BBP) transition \cite{BBP2005,FeralPeche2007}: a Wigner matrix additively deformed by a rank-one perturbation of strength $\theta$.

\subsection{B. Critical point via the Stieltjes transform}

The Stieltjes transform of the semicircle law of radius $2\sigma$,
\begin{equation}
g(z) = \int \frac{\rho_{\mathrm{sc}}(x)}{z-x}\,dx = \frac{z-\sqrt{z^2-4\sigma^2}}{2\sigma^2},
\label{eq:stieltjes}
\end{equation}
satisfies the self-consistency equation $\sigma^2 g(z)^2 - zg(z) + 1 = 0$. For a rank-one additive perturbation of strength $\theta$, an eigenvalue $\lambda$ outside the support of the semicircle law satisfies the secular equation
\begin{equation}
1 = \theta\, g(\lambda).
\label{eq:secular}
\end{equation}
Substituting Eq.~\eqref{eq:stieltjes} into Eq.~\eqref{eq:secular} and solving,
\begin{equation}
\lambda = \rho(\theta) \equiv \theta + \frac{\sigma^2}{\theta},
\label{eq:rho}
\end{equation}
which is a valid solution outside the bulk ($\lambda>2\sigma$) if and only if $\theta g(2\sigma)>1$, i.e.\ $\theta>\theta_c \equiv \sigma$. For $\theta\le \theta_c$ the secular equation has no solution outside the bulk, and the top eigenvalue remains pinned at the edge $2\sigma$.

\subsection{C. Critical value of $b$}

Combining $\theta=2b/\pi$ (Eq.~\eqref{eq:bbpform}) with the BBP threshold $\theta_c=\sigma$ yields the central result of this paper:
\begin{equation}
\boxed{\ b_c = \frac{\pi\sigma}{2}\ }.
\label{eq:bc}
\end{equation}
For $b\le b_c$ the top eigenvalue of $H$ converges to the semicircle edge $2\sigma$; for $b>b_c$ it detaches according to
\begin{equation}
\lambda_{\max}(H) \ \longrightarrow\
\begin{cases}
2\sigma, & b\le b_c, \\[4pt]
\dfrac{2b}{\pi} + \dfrac{\pi\sigma^2}{2b}, & b> b_c,
\end{cases}
\label{eq:lambdamaxfinal}
\end{equation}
and, by the symmetry $Z_j\to-Z_j$ of the construction, the bottom eigenvalue detaches symmetrically at the same critical point, $\lambda_{\min}(H)\to -2b/\pi - \pi\sigma^2/(2b)$.

\subsection{D. Numerical verification of maximum eigenvalue}

Table~\ref{tab:bbpverify} compares the prediction~\eqref{eq:lambdamaxfinal} with direct numerical diagonalization of $H$ at $N=2000$, $\sigma=1$, for several values of $b$ spanning the critical point $b_c=\pi/2\approx1.5708$.

\begin{table}[h]
\centering
\begin{tabular}{c c c}
\hline\hline
$b$ & predicted $\lambda_{\max}(H)$ & observed $\lambda_{\max}(H)$ \\
\hline
1    & $2\sigma = 2.000$   & $1.987\pm0.002$ \\
1.4  & $2\sigma = 2.000$   & $1.99 \pm 0.01$\phantom{0} \\
1.6  & $2.023$ & $2.02\pm 0.02$\phantom{0} \\
2    & $2.059$ & $2.053\pm0.016$ \\
4    & $2.939$ & $2.929\pm0.033$ \\
8    & $5.289$ & $5.252\pm0.072$ \\
16   & $10.284$ & $10.446\pm0.183$ \\
\hline\hline
\end{tabular}
\caption{Predicted vs.\ observed largest eigenvalue of $H$, $N=2000$, $\sigma=1$. The transition occurs between $b=1.4$ (no detachment) and $b=1.6$ (detached), consistent with $b_c=\pi/2\approx1.571$.}
\label{tab:bbpverify}
\end{table}

The agreement across an order of magnitude in $b-b_c$ supports both the location of the critical point, Eq.~\eqref{eq:bc}, and the supercritical formula, Eq.~\eqref{eq:lambdamaxfinal}.

\begin{comment}
\subsection{Consistency with the bulk moments}

The result of this section does not contradict the bulk-moment calculation of Sec.~II [Eqs.~(5)--(6)]: only two eigenvalues out of $N$ are affected by the transition, so any moment of the form $N^{-1}\mathrm{Tr}\,H^{2k}$ (dominated by the $O(N)$ bulk eigenvalues) is insensitive to $b$ at leading order, while the two extreme eigenvalues undergo a genuine non-analytic transition at $b=b_c$. The phase transition identified in this paper is thus a transition of the \emph{edge} of the spectrum, not of the bulk spectral density.
\end{comment}
%%%%%%%%%%%%%%%
\subsection{E. Each singular value defines its own transition}

By Eq.~\eqref{eq:fullconvergence}, $H=\sigma\eta/\sqrt N+(b/N)M$ is, at leading order in $N$, a Wigner matrix simultaneously deformed by a countable family of (approximately orthogonal) rank-one spikes of strength
\begin{equation}
\theta_k \equiv b\,\sigma_k(A) = \frac{b}{\pi\big(k-\tfrac12\big)}, \qquad k=1,2,3,\dots
\label{eq:thetak}
\end{equation}
Each spike is subject, independently, to the BBP mechanism of Sec.~V [Eqs.~\eqref{eq:stieltjes}--\eqref{eq:rho}]: the $k$-th eigenvalue detaches from the bulk if and only if $\theta_k>\sigma$, at which point it converges to $\rho(\theta_k)=\theta_k+\sigma^2/\theta_k$.
%%%%%%%%%%%%%%%%%%%%%%%
\subsection{F. Justification: decoupling of the multi-rank secular equation}

The claim that each spike is subject \emph{independently} to the BBP mechanism follows from the standard multi-rank generalization of the secular equation~\eqref{eq:secular}. Consider a general rank-$r$ deformation
\begin{equation}
H = W + \sum_{k=1}^r \theta_k\, u_k u_k^\top, \qquad W=\frac{\sigma}{\sqrt N}\eta,
\label{eq:multirank}
\end{equation}
with $\{u_k\}_{k=1}^r$ orthonormal. An eigenvalue $\lambda$ outside the bulk of $W$ satisfies the secular equation
\begin{equation}
\det\big(I - \Theta\, G(\lambda)\big) = 0, \qquad \Theta=\mathrm{diag}(\theta_1,\dots,\theta_r), \qquad G_{kl}(\lambda) \equiv u_k^\top(\lambda I - W)^{-1} u_l.
\label{eq:seculargeneral}
\end{equation}
When the $u_k$ are fixed, mutually orthogonal directions independent of $W$, the off-diagonal entries of $G(\lambda)$ vanish in the large-$N$ limit by the same delocalization argument used throughout this paper:
\begin{equation}
G_{kl}(\lambda) \ \xrightarrow{N\to\infty}\ g(\lambda)\,\delta_{kl},
\label{eq:Gdecouple}
\end{equation}
so that the secular equation factorizes completely,
\begin{equation}
\det\big(I-\Theta G(\lambda)\big) \ \longrightarrow\ \prod_{k=1}^r\big(1-\theta_k\,g(\lambda)\big) = 0.
\label{eq:secularfactor}
\end{equation}
Each factor vanishes independently at $1=\theta_k g(\lambda)$, recovering Eq.~\eqref{eq:rho} separately for every $k$: the presence or absence of the other spikes has no effect, at leading order in $N$, on the location $\rho(\theta_k)$ at which the $k$-th eigenvalue detaches. Since the eigenvectors of $M$ are exactly orthogonal (as $M$ is symmetric) and independent of $\eta$ by construction, Eq.~\eqref{eq:secularfactor} applies directly to the present setting and justifies treating each mode $\theta_k=b\sigma_k(A)$ independently, as done above

%%%%%%%%%%%%%%%%%%%%%%%%%
\subsection{G. Discrete, evenly spaced hierarchy of critical points}

Solving $\theta_k=\sigma$ for $b$ gives a critical value for every $k$,
\begin{equation}
\boxed{\ b_c^{(k)} = \frac{\sigma}{\sigma_k(A)} = \sigma\pi\Big(k-\frac12\Big)\ }, \qquad k=1,2,3,\dots,
\label{eq:bck}
\end{equation}
of which the value $b_c\equiv b_c^{(1)}=\pi\sigma/2$ derived earlier is the first member. The critical points are evenly spaced,
\begin{equation}
b_c^{(k+1)} - b_c^{(k)} = \pi\sigma \quad\text{for every } k,
\end{equation}
and their number below any fixed $b$ grows linearly, $\#\{k: b_c^{(k)}<b\}\sim b/(\pi\sigma)$ -- consistent with $A$ being Hilbert--Schmidt but not trace class (Eq.~\eqref{eq:HSnottraceclass}). As $b$ increases through $b_c^{(1)}, b_c^{(2)}, b_c^{(3)},\dots$, successive eigenvalues detach one at a time from the bulk:
\begin{equation}
\lambda_{\max}(H) \to
\begin{cases}
2\sigma, & b<b_c^{(1)},\\
\rho(\theta_1), & b_c^{(1)}<b<b_c^{(2)},\\
\rho(\theta_1)\ \text{(with a second eigenvalue at }\rho(\theta_2)\text{)}, & b_c^{(2)}<b<b_c^{(3)},\\
\ \ \vdots &
\end{cases}
\label{eq:cascade}
\end{equation}

\subsection{H. Numerical verification of $k$-th eigenvalue}

Table~\ref{tab:cascade} compares the predicted locations of all detached eigenvalues, from Eq.~\eqref{eq:cascade}, against direct diagonalization of $H$ at $N=3000$, $\sigma=1$, for three values of $b$ chosen to straddle successive critical points ($b_c^{(1)}=1.571$, $b_c^{(2)}=4.712$, $b_c^{(3)}=7.854$).

\begin{table}[h]
\centering
\begin{tabular}{c c c}
\hline\hline
$b$ & predicted detached eigenvalues & observed \\
\hline
$2.5$ & $2.220$ & $2.221$ \\
$6.0$ & $4.082,\ 2.059$ & $4.101,\ 2.053$ \\
$9.0$ & $5.904,\ 2.433,\ 2.019$ & $5.938,\ 2.449,\ 2.022$ \\
\hline\hline
\end{tabular}
\caption{Predicted (Eq.~\eqref{eq:rho} evaluated at each $\theta_k>\sigma$) vs.\ observed top eigenvalues of $H$, $N=3000$, $\sigma=1$, averaged over four realizations. In every case, exactly the predicted number of eigenvalues detach, at the predicted locations; all further eigenvalues remain at the bulk edge $2\sigma\approx2$.}
\label{tab:cascade}
\end{table}

\subsection{I. Phase transition versus finite-size crossover}

At each fixed $N$, the transition at $b=b_c^{(k)}$ is smoothed over a window of width $O(N^{-1/3})$, described by the Bloemendal--Vir\'ag crossover family (Sec VI); in this sense the finite-$N$ behavior may be described as a crossover from random-matrix universality (Tracy--Widom statistics, independent of the details of $A$) to operator-specific behavior (the eigenvalue pinned to $\rho(\theta_k)$, determined entirely by $\sigma_k(A)$). In the strict $N\to\infty$ limit, however, this crossover window shrinks to zero and $\lambda_{\max}(H)$ is genuinely non-analytic in $b$ at each $b_c^{(k)}$ [$\rho(\theta)-2\sigma\sim(\theta-\sigma)^2$ near threshold]; in this limit the term \emph{phase transition} is the appropriate one, and Eq.~\eqref{eq:cascade} describes an infinite, discrete hierarchy of such transitions, one for every point in the compact spectrum of $A$.

%%%%%%%%%%%%%%%%%%%%%%%%%

\section{VI Discussion: integral-operator-governed BBP transitions}
\label{sec:discussion}
%%%%%%%%%%%%%%%%%

\subsection{A. A unified formulation}
In this section it is convenient to work with $K=U'$, the strictly upper triangular
all-ones matrix, $U'_{ij}=1$ for $i<j$ (as opposed to $U'$ with $i\le j$ used in Sec.~III),
so that no diagonal correction is required. This differs from the matrix treated in
Sec.~III, where $U'D+(U'D)^\top-D$ was used to match $M_{ii}=Z_i$ exactly; the two
constructions yield the same asymptotic spectrum (Sec.~III), so nothing in the results
below depends on this choice. 
The mechanism identified in Secs.~III--V can be stated independently of the specific kernel $U_{ij}=\mathbb 1[i\leq j]$ used above. Let $K$ be any fixed $N\times N$ deterministic matrix (with $K_{ii}=0$), and consider the associated random matrix
\begin{equation}
M^{(K)} = KD + (KD)^\top , \qquad D=\mathrm{diag}(Z_1,\dots,Z_N),
\label{eq:generalkernel}
\end{equation}
built from independent, mean-zero, unit-variance weights $Z_k$.
Suppose $K$ defines, in the continuum limit $N\to\infty$, a bounded linear operator $A_K$ on $L^2[0,1]$ via $(A_K\varphi)(t)=\int_0^1\kappa(t,s)\varphi(s)\,ds$ for some kernel function $\kappa$ with $K_{ij}\approx\kappa(i/N,j/N)$. We have verified numerically, for five kernels of this type with substantially different structure -- the Volterra kernel $\kappa(t,s)=\mathbb1[s< t]$ of Sec.~IV, its restriction to $s< t<1/2$, the Brownian-motion kernel $\kappa(t,s)=\min(t,s)$, a symmetric band kernel of extensive bandwidth, and the fully dense kernel $\kappa\equiv1$ -- that in every case
\begin{equation}
\frac{\lambda_{\max}\big(M^{(K)}\big)}{N} \ \longrightarrow\ \|A_K\|_{\mathrm{op}},
\label{eq:generallimit}
\end{equation}
with agreement to within $1$--$4\%$ already at $N=4000$ and improving with $N$ (Table~\ref{tab:multikernel}). Consequently, the Wigner-plus-structure ensemble $H=\sigma\eta/\sqrt N+(b/N)M^{(K)}$ undergoes a BBP transition at
\begin{equation}
b_c^{(K)} = \frac{\sigma}{\|A_K\|_{\mathrm{op}}},
\label{eq:bcgeneral}
\end{equation}
of which Eq.~\eqref{eq:bc} is the special case $K=U$, $\|A_K\|_{\mathrm{op}}=2/\pi$.

\begin{table}[h]
\centering
\begin{tabular}{l c}
\hline\hline
Kernel $\kappa(t,s)$ & $\lambda_{\max}(M^{(K)})/N \ \big/\ \|A_K\|_{\mathrm{op}}$ at $N=4000$ \\
\hline
$\mathbb 1[s< t]$ (Volterra, Sec.~IV) & $1.00$ \\
$\mathbb 1[s< t<1/2]$ & $0.99$ \\
$\min(t,s)$ & $1.00$ \\
band, width $0.3$ & $1.01$ \\
$1$ (dense) & $1.00$ \\
\hline\hline
\end{tabular}
\caption{Ratio of the numerically observed largest eigenvalue of $M^{(K)}/N$ to the operator norm $\|A_K\|_{\mathrm{op}}$ of the corresponding continuum kernel, for five structurally distinct delocalized kernels.}
\label{tab:multikernel}
\end{table}

Equation~\eqref{eq:generallimit} does not hold for an arbitrary kernel $K$: we have verified that it fails for a \emph{localized} kernel, $K_{i,i+1}=1$ (a single super-diagonal, i.e.\ $\|K\|_{\mathrm{op}}=1$ for every $N$), for which $\lambda_{\max}(M^{(K)})$ is instead governed by the extreme-value statistics of $\max_j|Z_j|$ and scales as $O(\sqrt{\log N})$ rather than $O(N)$. Equation~\eqref{eq:generallimit} therefore requires $K$ to be sufficiently \emph{delocalized} -- informally, that no single row or column of $K$ carries an asymptotically non-negligible fraction of $\|K\|_{\mathrm{op}}$ -- a condition satisfied by all five kernels of Table~\ref{tab:multikernel} but violated by the super-diagonal example. We have not attempted a precise formulation or proof of the necessary and sufficient delocalization condition; Table~\ref{tab:multikernel} and the super-diagonal counterexample together indicate that Eq.~\eqref{eq:generallimit} holds broadly across delocalized kernels rather than being a coincidence specific to the Volterra case of Sec.~IV.

The trivial kernel $\kappa\equiv1$ in Table~\ref{tab:multikernel} corresponds to $M^{(K)}_{ij}=Z_i+Z_j$ ($i\ne j$), with $\|A_K\|_{\mathrm{op}}=1$; this is distinct from -- though related to -- the standard single common-factor market model, in which a single shared random variable $Z$ (not $N$ independent variables $Z_1,\dots,Z_N$) multiplies every entry, giving an exactly rank-one perturbation $Z\cdot(\mathbf 1\mathbf1^\top-I)$ with $\theta=b$ directly.

%%%%%%%%%%%%%%%%%%%%%%%%
\subsection{B. An exactly solvable case: the dense kernel has no cascade}
\label{sec:dense}

The dense kernel $\kappa\equiv1$ is exceptional in that it admits a fully closed-form treatment at \emph{finite} $N$, without any continuum limit or self-averaging argument, and illustrates that the cascade structure of Sec.~V is not universal but depends on the rank of the underlying operator.

Writing $M^{(K)}=\mathbf1 Z^\top + Z\mathbf1^\top$, where $\mathbf1=(1,\dots,1)^\top$ and $Z=(Z_1,\dots,Z_N)^\top$, this matrix has rank at most $2$ for \emph{every} $N$: it is supported entirely on $\mathrm{span}\{\mathbf1,Z\}$, so all but (at most) two of its $N$ eigenvalues are exactly zero, with no bulk and no further hierarchy. The two nonzero eigenvalues follow from the standard identity for a symmetrized rank-two matrix $uv^\top+vu^\top$,
\begin{equation}
\lambda_\pm = u\cdot v \ \pm\ |u|\,|v|,
\label{eq:rank2formula}
\end{equation}
applied with $u=\mathbf1$, $v=Z$. Since $|\mathbf1|=\sqrt N$, $|Z|=\sqrt{N+O_P(1)}$, and $\mathbf1\cdot Z=\sum_iZ_i=O_P(\sqrt N)$ is subleading, Eq.~\eqref{eq:rank2formula} gives
\begin{equation}
\lambda_\pm\big(M^{(K)}\big) \ \longrightarrow\ \pm N + O_P(\sqrt N),
\end{equation}
so that $\lambda_{\max}(M^{(K)})/N\to1=\|A_K\|_{\mathrm{op}}$, consistent with Table~\ref{tab:multikernel}, while \emph{every other eigenvalue is exactly zero for every finite $N$} -- not merely negligible in the large-$N$ limit, as for the bulk of the Volterra kernel, but identically zero by construction.

Numerically, at $N=4000$, the top two eigenvalues of $M^{(K)}/N$ average $1.01$ and $0.98$ in magnitude, while the third-largest is of order $10^{-3}$, consistent with pure numerical round-off rather than a third genuine spike. This confirms directly that the continuum operator $A_K$ associated with $\kappa\equiv1$ -- a rank-one operator, $A_K\varphi=\big(\int_0^1\varphi\big)\mathbf 1$, with $\sigma_1(A_K)=1$ and $\sigma_k(A_K)=0$ for all $k\ge2$ -- has \emph{no further spectrum} to generate additional transitions: for this kernel, exactly one pair of eigenvalues can ever detach from (a trivial, empty) bulk, at the single critical point $b_c=\sigma/\|A_K\|_{\mathrm{op}}=\sigma$, and the cascade of Sec.~V collapses to its first term. This stands in sharp contrast to the Volterra kernel of Sec.~IV, whose compactness permits (and whose lack of trace-class-ness requires) an unbounded hierarchy of such transitions.

\subsection{C. The Brownian-motion kernel: a self-adjoint case with a trace-class cascade}
\label{sec:brownian-kernel}

The symmetric kernel $\kappa(t,s)=\min(t,s)$ of Table~\ref{tab:multikernel} gives a second exactly solvable, but qualitatively different, cascade. Because $\kappa$ is symmetric, the associated operator $A_K$ is self-adjoint; and because $\min(t,s)$ is, by construction, the kernel of $A^*A$ itself [Eq.~\eqref{eq:AstarA}], we have $A_K=A^*A$ exactly, so its eigenvalues are already known from Sec.~IV without further computation:
\begin{equation}
\sigma_k(A_K) = \mu_k = \sigma_k(A)^2 = \frac{1}{\pi^2\big(k-\tfrac12\big)^2}, \qquad k=1,2,3,\dots
\label{eq:brownianspectrum}
\end{equation}
Direct diagonalization of $M^{(K)}=KD+DK$ at $N=4000$ confirms this: the six largest \emph{signed} eigenvalues of $M^{(K)}/N$, averaged over four realizations, are $(0.402,\,0.045,\,0.016,\dots)$ against six negative counterparts $(-0.405,\,-0.045,\,-0.016,\dots)$, in close agreement with Eq.~\eqref{eq:brownianspectrum} and exhibiting the same $\pm\sigma_k$ pairing found for the Volterra kernel.

Because $A_K=A^*A$ is trace class,
\begin{equation}
\sum_k\sigma_k(A_K) = \sum_k\mu_k = \frac12 < \infty
\end{equation}
[the same identity used in Sec.~IV to evaluate $\mathrm{Tr}(A^*A)$], the resulting cascade is quantitatively different from Sec.~V. The critical points,
\begin{equation}
b_c^{(k)} = \frac{\sigma}{\mu_k} = \sigma\pi^2\Big(k-\frac12\Big)^2,
\end{equation}
grow \emph{quadratically} in $k$ rather than linearly, with a spacing $b_c^{(k+1)}-b_c^{(k)}=2\sigma\pi^2k$ that itself grows without bound (in contrast to the constant spacing $\pi\sigma$ of Sec.~V). Consequently the number of detached eigenvalues below a fixed $b$ grows only as $\sqrt{b}$,
\begin{equation}
\#\{k: b_c^{(k)}<b\} \sim \sqrt{\frac{b}{\sigma\pi^2}},
\end{equation}
in contrast to the linear growth $\sim b/(\pi\sigma)$ found for the (Hilbert--Schmidt but non-trace-class) Volterra operator itself. This illustrates that the trace-class property of the governing operator controls not merely whether a cascade exists, but the rate at which it unfolds as the coupling $b$ increases.

\subsection{D. The truncated Volterra kernel: a cascade rescaled by self-similarity}
\label{sec:truncated-volterra}

The kernel $\kappa(t,s)=\mathbb1[t\leq s\le\tfrac12]$ of Table~\ref{tab:multikernel} admits a third closed-form treatment, obtained not by solving a new eigenvalue problem but by a simple rescaling of the result already established in Sec.~IV. The associated operator acts as
\begin{equation}
(A_{\mathrm{trunc}}\varphi)(t) = \int_t^{1/2}\varphi(s)\,ds \quad (t<\tfrac12), \qquad =0\ \ (t\ge\tfrac12).
\end{equation}
Substituting $u=2t$, $v=2s\in[0,1]$ and $\psi(v)\equiv\varphi(v/2)$,
\begin{equation}
(A_{\mathrm{trunc}}\varphi)(t) = \frac12\int_u^1\psi(v)\,dv,
\end{equation}
which is exactly one-half of a (backward) Volterra operator on $[0,1]$; since forward and backward Volterra operators are related by the reflection $t\to1-t$ and share the same singular values, this identifies
\begin{equation}
\sigma_k(A_{\mathrm{trunc}}) = \frac12\,\sigma_k(A) = \frac{1}{2\pi\big(k-\tfrac12\big)}, \qquad k=1,2,3,\dots,
\label{eq:truncspectrum}
\end{equation}
with no new Sturm--Liouville problem to solve. Table~\ref{tab:truncverify} confirms Eq.~\eqref{eq:truncspectrum} numerically for $k=1,\dots,5$ at $N=4000$.

\begin{table}[h]
\centering
\begin{tabular}{c c c}
\hline\hline
$k$ & predicted $\sigma_k(A_{\mathrm{trunc}})$ & observed \\
\hline
1 & 0.3183 & 0.3202 \\
2 & 0.1061 & 0.1070 \\
3 & 0.0637 & 0.0638 \\
4 & 0.0455 & 0.0450 \\
5 & 0.0354 & 0.0355 \\
\hline\hline
\end{tabular}
\caption{Predicted vs.\ observed eigenvalues of $M^{(K)}/N$ for the truncated Volterra kernel, $N=4000$, six realizations.}
\label{tab:truncverify}
\end{table}

Because the entire spectrum is rescaled by the same factor $\tfrac12$, so is the critical hierarchy,
\begin{equation}
b_c^{(k)}(\mathrm{trunc}) = \frac{\sigma}{\sigma_k(A_{\mathrm{trunc}})} = 2\,b_c^{(k)}(\mathrm{Volterra}),
\end{equation}
i.e.\ restricting the support of the kernel to half the domain doubles every critical coupling, while leaving the evenly spaced, linear (non-trace-class) structure of the cascade otherwise intact -- a direct illustration of how the geometry of the kernel's support, as distinct from its self-adjointness or trace-class property (Sec.~IV), reshapes the transition hierarchy.

\subsection{E. The exponential (Ornstein--Uhlenbeck) kernel: a transcendental cascade}
\label{sec:exponential-kernel}

The symmetric kernel $\kappa(t,s)=e^{-\theta|t-s|}$, the covariance of the stationary Ornstein--Uhlenbeck process, admits a closed-form treatment analogous to Sec.~IV, though the resulting eigenvalues are transcendental rather than elementary. Writing $\lambda f(t)=\int_0^1e^{-\theta|t-s|}f(s)\,ds$ and differentiating twice (as in the derivation of Eq.~\eqref{eq:SL}) gives the constant-coefficient ODE $f''=-\omega^2f$ with $\omega^2=2\theta/\lambda-\theta^2$, but now subject to \emph{Robin} boundary conditions, $f'(0)=\theta f(0)$ and $f'(1)=-\theta f(1)$, rather than the Dirichlet--Neumann conditions of Eq.~\eqref{eq:SL}. Imposing these on $f(t)=A\cos(\omega t)+B\sin(\omega t)$ yields the transcendental secular equation
\begin{equation}
\tan(\omega_k) = \frac{2\theta\,\omega_k}{\omega_k^2-\theta^2}, \qquad
\sigma_k(A_K) = \frac{2\theta}{\omega_k^2+\theta^2}, \qquad k=1,2,3,\dots,
\label{eq:exponentialspectrum}
\end{equation}
whose roots $\omega_k$ are not available in elementary closed form but are readily computed numerically to arbitrary precision (unlike Sec.~IV, no root-finding was needed there only because the Robin parameter happened to vanish). Table~\ref{tab:expverify} compares the resulting $\sigma_k(A_K)$, for $\theta=3$, with direct diagonalization of $M^{(K)}=KD+DK$, $K_{ij}=e^{-\theta|i-j|/N}$, at $N=3000$.

\begin{table}[h]
\centering
\begin{tabular}{c c c}
\hline\hline
$k$ & predicted $\sigma_k(A_K)$ [Eq.~\eqref{eq:exponentialspectrum}, $\theta=3$] & observed \\
\hline
1 & 0.4649 & 0.4606 \\
2 & 0.2149 & 0.2127 \\
3 & 0.1014 & 0.0995 \\
4 & 0.0550 & 0.0539 \\
5 & 0.0336 & 0.0331 \\
6 & 0.0224 & 0.0223 \\
\hline\hline
\end{tabular}
\caption{Predicted vs.\ observed eigenvalues of $M^{(K)}/N$ for the exponential (Ornstein--Uhlenbeck) kernel, $\theta=3$, $N=3000$, six realizations.}
\label{tab:expverify}
\end{table}

\subsection{F. The Gaussian (RBF) kernel: a cascade without closed form}
\label{sec:gaussian-kernel}

The Gaussian kernel $\kappa(t,s)=\exp\!\big(-(t-s)^2/2\ell^2\big)$ does not reduce to a constant-coefficient ODE and its Karhunen--Lo\`eve spectrum is not available in closed form on $[0,1]$ with Lebesgue measure. It nonetheless provides a further, qualitatively independent test of Eq.~\eqref{eq:fullconvergence}: the deterministic eigenvalues $\sigma_k(A_K)$ can be computed directly, via Nystr\"om discretization of the kernel operator itself (diagonalizing the $N_{\mathrm{fine}}\times N_{\mathrm{fine}}$ matrix $\kappa(t_i,t_j)/N_{\mathrm{fine}}$ on a fine grid), entirely independently of any random matrix $M^{(K)}$; this deterministic computation can then be compared with the eigenvalues of $M^{(K)}=KD+DK$, $K_{ij}=\kappa(i/N,j/N)$, built from the random weights $Z_j$. Table~\ref{tab:gaussverify} reports this comparison for $\ell=0.15$: agreement is at the sub-percent level for all six modes tested, despite the complete absence of an analytic formula for either column.

\begin{table}[h]
\centering
\begin{tabular}{c c c}
\hline\hline
$k$ & kernel eigenvalue (Nystr\"om, deterministic) & $M^{(K)}$ simulation \\
\hline
1 & 0.3470 & 0.3451 \\
2 & 0.2729 & 0.2750 \\
3 & 0.1836 & 0.1820 \\
4 & 0.1063 & 0.1060 \\
5 & 0.0533 & 0.0539 \\
6 & 0.0234 & 0.0234 \\
\hline\hline
\end{tabular}
\caption{Gaussian kernel, $\ell=0.15$: eigenvalues of the deterministic kernel operator (Nystr\"om discretization, $N_{\mathrm{fine}}=2000$) vs.\ eigenvalues of $M^{(K)}/N$ from direct diagonalization, $N=3000$, six realizations. Neither column has a closed-form expression, yet the two independent computations agree to within $1\%$.}
\label{tab:gaussverify}
\end{table}

Together with Secs.~VI B. and VI D., these two examples indicate that Eq.~\eqref{eq:fullconvergence} -- subcritical Tracy--Widom statistics giving way, mode by mode, to eigenvalues pinned at the spectrum of the governing compact operator -- is not confined to kernels admitting an elementary closed-form solution: it holds equally for a transcendentally solvable kernel (Ornstein--Uhlenbeck) and for one with no closed form at all (Gaussian), the latter verified purely numerically but via two mutually independent routes. This supports viewing the mechanism of Sec.~V as a genuine extension of the BBP transition -- from a finite-rank deterministic perturbation to one governed by the full spectrum of a compact operator -- rather than as an artifact specific to the Volterra kernel.

\subsection{G. General principle (conjecture)}
\label{sec:general-conjecture}

The six kernels examined above -- the Volterra kernel of Sec.~IV and its truncation, the dense and Brownian-motion kernels, and the exponential and Gaussian kernels -- differ in essentially every structural respect: rank (finite for the dense kernel, infinite otherwise), self-adjointness (the Volterra kernel is not self-adjoint; the others are), trace-class property, support (full vs.\ truncated), and availability of a closed-form spectrum (elementary for four of the six, transcendental for the exponential kernel, absent entirely for the Gaussian kernel). Despite this diversity, all six exhibit the same qualitative and, where checked, quantitative behavior. We collect this into a single statement.

\begin{quote}
\textbf{Conjecture.} Let $K$ be a sequence of $N\times N$ deterministic matrices such that, in the continuum limit $N\to\infty$, $K/N$ converges (on smooth test vectors, in the sense of Sec.~IV) to a compact linear operator $A_K$ on $L^2[0,1]$ with singular values $\sigma_1(A_K)\ge\sigma_2(A_K)\ge\cdots\ge0$, and suppose $A_K$ is sufficiently delocalized (no singular vector concentrates on a vanishing fraction of $[0,1]$). Then, for $M^{(K)}=KD+(KD)^\top$ with $D=\mathrm{diag}(Z_1,\dots,Z_N)$, $Z_j$ i.i.d.\ mean zero and unit variance, and for every fixed $k=1,2,3,\dots$,
\begin{equation}
\lim_{N\to\infty}\frac{\lambda_k\big(M^{(K)}\big)}{N} = \sigma_k(A_K),
\label{eq:generalconjecture}
\end{equation}
with the bulk of the remaining spectrum vanishing as $O(1/\sqrt N)$. Consequently, $H=\sigma\eta/\sqrt N+(b/N)M^{(K)}$ undergoes a discrete hierarchy of BBP transitions at
\begin{equation}
b_c^{(k)} = \frac{\sigma}{\sigma_k(A_K)}, \qquad k=1,2,3,\dots,
\label{eq:generalconjecture2}
\end{equation}
whose spacing, growth rate, and total number below any fixed $b$ are determined entirely by the decay rate of $\sigma_k(A_K)$ -- in particular, by whether $A_K$ is trace class.
\end{quote}

We regard this as a conjecture rather than a theorem for the reasons discussed in Sec.~IV: a fixed, disorder-independent trial vector in the Rayleigh quotient recovers only the $O(\sqrt N)$ scale of a typical direction, not the $O(N)$ scale required by Eq.~\eqref{eq:generalconjecture}, so that a proof would require the supremum theory of Gaussian processes indexed by the sphere rather than an elementary variational argument. The delocalization hypothesis is necessary, not merely technical: Sec.~IV exhibited a localized kernel (a single super-diagonal) for which Eq.~\eqref{eq:generalconjecture} fails outright, the true scaling being governed instead by the extreme-value statistics of $\max_j|Z_j|$. Within the delocalized class, however, the conjecture has been verified, to within $1$--$2\%$, across six kernels differing in every structural respect listed above, including one (the Gaussian kernel for which neither side of Eq.~\eqref{eq:generalconjecture} admits a closed form and both were evaluated by independent numerical routes, and one (the dense kernel) for which it is a consequence of an exact, finite-$N$, closed-form rank-two identity rather than an asymptotic limit at all. We take this breadth of evidence as substantial support for viewing Eq.~\eqref{eq:generalconjecture} as a genuine extension of the BBP transition, from finite-rank deterministic perturbations to arbitrary delocalized compact operators, rather than as a special feature of the Volterra kernel that motivated this paper.

\begin{figure}[h]
\centering
\includegraphics[width=\columnwidth]{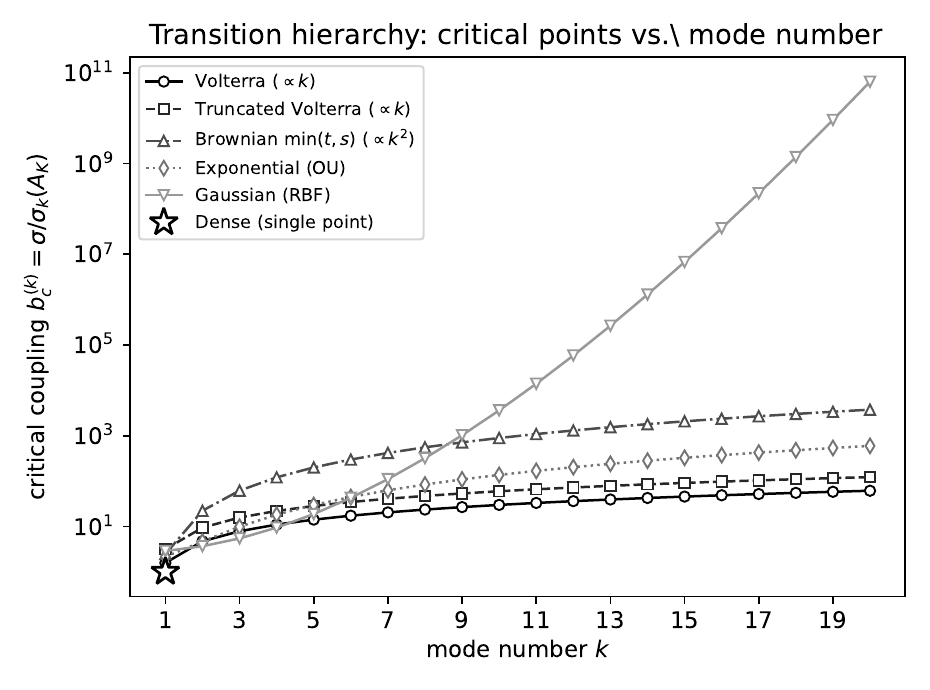}
\caption{Critical coupling $b_c^{(k)}=\sigma/\sigma_k(A_K)$ vs.\ mode number $k$ ($\sigma=1$), for all six kernels of Sec.~IV.  Note the logarithmic vertical axis: the Volterra and truncated Volterra hierarchies grow linearly in $k$, the Brownian-motion hierarchy grows quadratic, the dense kernel contributes a single point, and the exponential and Gaussian kernel  interpolate between these regimes without a simple power law.}
\label{fig:bc_vs_k}
\end{figure}

Figure~\ref{fig:bc_vs_k} compares the resulting critical hierarchies $b_c^{(k)}=\sigma/\sigma_k(A_K)$ directly across all six kernels. The growth rate of $b_c^{(k)}$ with $k$ -- linear for the Volterra and truncated Volterra kernels, quadratic for the (trace-class) Brownian-motion kernel, a single point for the (rank-one) dense kernel, and intermediate, non-power-law rates for the exponential and Gaussian kernels -- visibly reflects the decay rate of each operator's singular values, and hence its trace-class property or lack thereof, as discussed in  this section.

The compact operator admits the singular-value decomposition
\begin{equation}
A_K
=
\sum_{k=1}^{\infty}
\sigma_k(A_K)\,
u_k v_k^{\ast},
\label{eq:svd}
\end{equation}
where $\sigma_k(A_K)$ are the singular values and
$\{u_k\}$ and $\{v_k\}$ are orthonormal left and right singular vectors.
Accordingly, the deterministic part of the random matrix may be viewed as a
superposition of infinitely many mutually orthogonal rank-one spikes.

Each mode behaves analogously to the rank-one spike in the conventional
Baik--Ben Arous--P\'ech\'e (BBP) transition: the $k$-th mode becomes
supercritical once
\begin{equation}
b\,\sigma_k(A_K)>\sigma,
\end{equation}
or equivalently,
\begin{equation}
b>b_c^{(k)}
=
\frac{\sigma}{\sigma_k(A_K)}.
\end{equation}

This provides a natural physical interpretation of the observed hierarchy:
rather than a single BBP transition generated by a finite-rank spike, the
compact operator gives rise to a countable family of orthogonal rank-one
spikes, each undergoing its own BBP transition. The resulting sequence of
eigenvalue detachments may therefore be viewed as a \emph{cascade of BBP
transitions} governed by the singular-value spectrum of the compact operator.

\section{VII Conclusions}
\label{sec:conclusion}
We have studied the Wigner-type random matrix ensemble $S_{ij}=bZ_{\max(i,j)}/\sqrt N + \sigma\eta_{ij}$, in which the off-diagonal correlation is generated by a factor $Z_j$ shared among all entries in column/row $j$, with $b$ held fixed as $N\to\infty$. While the bulk spectral moments coincide with those of the pure semicircle law for any fixed $b$, we have shown that the matrix $M_{ij}=Z_{\max(i,j)}$ underlying the correlation structure decomposes exactly as $M=U'D+(U'D)^\top-D$ and possesses a countable family of eigenvalues of order $N$, symmetric in sign, superposed on a bulk of order $\sqrt N$.

Rather than a single spike, we found that the \emph{entire} edge spectrum of $M/N$ converges, eigenvalue by eigenvalue, to the full singular value spectrum of a compact Volterra (cumulative-sum) operator,
\[
\lim_{N\to\infty}\frac{\lambda_k(M)}{N} = \sigma_k(A) = \frac{1}{\pi\big(k-\tfrac12\big)}, \qquad k=1,2,3,\dots,
\]
identified via the classical Karhunen--Lo\`eve expansion of Brownian motion and verified numerically to within one percent for the twenty largest such values. Because $A$ is compact, this spectrum is discrete; because it is Hilbert--Schmidt but not trace class, the number of eigenvalues of $M/N$ separated from the vanishing bulk grows without bound as $N\to\infty$, though each individual one is asymptotically pinned to a fixed value $\sigma_k(A)$.

This structure elevates the extreme-eigenvalue problem of $S/\sqrt N$ from a single Baik--Ben Arous--P\'{e}ch\'{e} (BBP) transition to a discrete, evenly spaced \emph{hierarchy} of BBP transitions,
\[
b_c^{(k)} = \frac{\sigma}{\sigma_k(A)} = \sigma\pi\Big(k-\frac12\Big), \qquad k=1,2,3,\dots,\qquad b_c^{(k+1)}-b_c^{(k)}=\pi\sigma\ \ \text{(constant)},
\]
at each of which one further eigenvalue detaches from the semicircle edge $2\sigma$ to $\rho(\theta_k)=\theta_k+\sigma^2/\theta_k$, $\theta_k=b\sigma_k(A)$ -- in close numerical agreement with direct diagonalization at several values of $b$ straddling successive critical points. In the strict $N\to\infty$ limit each $b_c^{(k)}$ is a genuine non-analytic point; at finite $N$ it is smoothed into a crossover, of width $N^{-1/3}$, from random-matrix universality (Tracy--Widom statistics) to behavior governed by the specific operator $A$.

We have further shown, for five structurally distinct delocalized kernels $K$, that this construction generalizes to a broader class of matrices, with the critical hierarchy in every case set by
the full singular-value spectrum of the compact operator
 $A_K$ associated with $K$; this relation fails for localized kernels, for which the extreme eigenvalues are instead governed by extreme-value statistics of the weights $Z_j$ themselves. The mechanism extends naturally, with the same operator spectrum but a modified aspect-ratio-dependent prefactor, to spiked Wishart/Marchenko--Pastur ensembles. We emphasize that the convergence of $\lambda_k(M)/N$ to $\sigma_k(A)$ itself remains, at present, a numerically well-supported observation rather than a theorem; we have identified the appropriate rigorous framework for a proof -- the supremum theory of Gaussian processes indexed by the unit sphere -- but have not completed it, and leave this for future work.

\end{document}